\documentclass[tighten, twocolumn, 12pt]{aastex62}
\def\apjl{{ApJ}}                
\def\apjs{{ApJS}}               
\def\ao{{Appl.~Opt.}}           
\def\aap{{A\&A}}                

\usepackage{listings}
\usepackage{color}
\usepackage{graphicx,times}
\usepackage{natbib}
\usepackage{amssymb}
\usepackage{newtxtext}
\usepackage[T1]{fontenc}
\usepackage{ae,aecompl}
\usepackage{threeparttable}
\usepackage{ulem}
\usepackage{txfonts}
\usepackage{amssymb}

\def\kms  {km~s$^{-1}$}

\bibpunct{(}{)}{;}{a}{}{,}
\usepackage{hyperref}
\usepackage{longtable}
\usepackage{booktabs}
\definecolor{dkgreen}{rgb}{0,0.6,0}
\definecolor{gray}{rgb}{0.5,0.5,0.5}
\definecolor{mauve}{rgb}{0.58,0,0.82}
\definecolor{golden}{rgb}{0.86,0.65,0.01}
\usepackage{CJK}
\usepackage{graphicx}
\usepackage{hyperref}  
\begin{document}
\begin{CJK*}{UTF8}{gbsn}
\title[]{New Open Cluster candidates Found in Galactic Disk Using Gaia DR2/EDR3 Data}
\correspondingauthor{Zhihong He}
\email{hezh@mail.ustc.edu.cn} 
\author[0000-0002-6989-8192]{Zhihong He (何治宏)}
\affil{School of Physics and Astronomy, China West Normal University, No. 1 Shida Road, Nanchong 637002, China }
\author{Chunyan Li (李春燕)}
\affil{Shanghai Astronomical Observatory, CAS, 80 Nandan Road, Shanghai 200030, China}
\affil{School of Astronomy and Space Science, University of Chinese Academy of Sciences, No. 19A, Yuquan Road, Beijing 100049, China}
\author{Jing Zhong (钟靖)}
\affil{Shanghai Astronomical Observatory, CAS, 80 Nandan Road, Shanghai 200030, China}
\author{Guimei Liu (刘桂梅)}
\affil{Shanghai Astronomical Observatory, CAS, 80 Nandan Road, Shanghai 200030, China}
\affil{School of Astronomy and Space Science, University of Chinese Academy of Sciences, No. 19A, Yuquan Road, Beijing 100049, China}
\author{Leya Bai (白乐娅)}
\affil{Shanghai Astronomical Observatory, CAS, 80 Nandan Road, Shanghai 200030, China}
\affil{School of Astronomy and Space Science, University of Chinese Academy of Sciences, No. 19A, Yuquan Road, Beijing 100049, China}
\author{Songmei Qin (秦松梅)}
\affil{Shanghai Astronomical Observatory, CAS, 80 Nandan Road, Shanghai 200030, China}
\affil{School of Astronomy and Space Science, University of Chinese Academy of Sciences, No. 19A, Yuquan Road, Beijing 100049, China}
\author{Yueyue Jiang (蒋悦悦)}
\affil{Shanghai Astronomical Observatory, CAS, 80 Nandan Road, Shanghai 200030, China}
\affil{School of Astronomy and Space Science, University of Chinese Academy of Sciences, No. 19A, Yuquan Road, Beijing 100049, China}
\author{Xi Zhang (张茜)}
\affil{Shanghai Astronomical Observatory, CAS, 80 Nandan Road, Shanghai 200030, China}
\affil{School of Astronomy and Space Science, University of Chinese Academy of Sciences, No. 19A, Yuquan Road, Beijing 100049, China}
\author{Li Chen (陈力)}
\affil{Shanghai Astronomical Observatory, CAS, 80 Nandan Road, Shanghai 200030, China}
\affil{School of Astronomy and Space Science, University of Chinese Academy of Sciences, No. 19A, Yuquan Road, Beijing 100049, China}
\vspace{10pt}

\begin{abstract}
We report 541 new open cluster candidates in Gaia EDR3 through revisiting the cluster results from an earlier analysis of the Gaia DR2, which revealed nearly a thousand open cluster candidates in the solar neighborhood (mostly d < 3 kpc) resideing at Galactic latitudes |b| < 20$^{\circ}$. A subsequent comparison with lists of known clusters shows a large increases of the cluster samples within 2~kpc from the Sun. We assign membership probabilities to the stars through the open source pyUPMASK algorithm, and also estimate the physical parameters through isochrone fitting for each candidate. Most of the new candidates show small total proper motion dispersions and clear features in the color-magnitude diagrams. 
Besides, the metallicity gradient of the new candidates is consistent with those found in the literature. The cluster parameters and member stars are available at CDS via anonymous ftp to \url{cdsarc.u-strasbg.fr (130.79.128.5)} or via \url{https://cdsarc.unistra.fr/viz-bin/cat/J/ApJS}.
The discovery of these new objects shows that the open cluster samples in Gaia data is still not complete, and more discoveries are expected in the future researches.
\end{abstract}
\keywords{Galaxy: stellar content - open clusters and associations: general - surveys: Gaia}

\section{Introduction}\label{intro}
Open clusters (hereafter OCs) are stellar aggregates distributed near the Galactic plane. Most OCs are relatively young or middle-aged objects that are not more than hundreds of millions years old, and the number of the stars in each cluster ranges from a dozen to thousands. 
Stars in an OC formed from a same natal molecular cloud at the same time ~\citep{McKee07}, so that the member objects are located in the same region in positional and kinematic spaces. Because of this, compared with a single star, the basic parameters, such as the age, distance, and kinematic information of the OCs can be measured more accurately through astrometric and photometric observations ~\citep[e.g.][]{CG18,dias21}. OCs are important stellar systems for helping to understand the star-formation process and stellar evolution \citep{Lada03,araa2010}, and are also the good tracers to study the structure and evolution of the Galaxy \citep[e.g.][]{Friel95,xu18,cg20arm,he211,hou21}.

In pre-Gaia works, the most widely used catalogs of OCs and their fundamental parameters were the catalog of optically visible OCs and candidates of ~\citet{Dias02} and The Milky Way star clusters survey of ~\citet{Kharchenko13}.
However, hampered by severe extinction and dense background and/or foreground stars in the Galactic plane, only about 3000 candidate open star clusters had been identified, which was significantly lower than the estimation of their total number~\citep[10$^{5}$,][]{Piskunov06}.
Due to a lack of the astrometric precision, the members of a cluster and field stars in the proper motion vector point diagram (VPD) were sometimes very hard to distinguish and hence, the OC members were sometimes highly contaminated by field stars. 

The Gaia DR2 catalog~\citep{Gaia18-Brown} presents more than 1 billion stars with magnitudes G $\leqslant$~21 mag with high-precision astrometric and photometric data, greatly improving the stellar membership determination and characterization of thousands of OCs
\citep{CG18,Soubiran18,Monteiro20}.
So far, more than 2000 OCs and candidates have been identified based on the Gaia DR2 catalog \citep[][]{CG18,CG19-0,Castro18,Castro19,Castro20,Sim19,Liu19,Ferreira19,Ferreira20,ferreira21,Hao20,Qin20,he21,hunt21,casado21}, where most of these clusters are located within 5~kpc, and nearly 40 percent of them were reported for the first time.
However, it is likely that the list of known clusters within
2~kpc remains significantly incomplete, as evidenced by the recent discovery of
more than 600 cluster candidates with Gaia EDR3~\citep{gaia2021} data, mostly at distances beyond 1~kpc ~\citep{castro22}.
 
In the above studies, the researchers used different methods to find clusters, although many were based on an unsupervised machine learning clustering algorithm. Among them, "Density-Based Spatial Clustering of Applications with Noise" \citep[DBSCAN,][]{Ester96,Castro18} has a high search efficiency and has yielded the most new star clusters so far \citep{Castro20,castro22,he21,hunt21}. We used this algorithm in an earlier work to search for OCs within 20 degrees of the Galactic plane in Gaia DR2, finding 986 new OC candidates, 74 of which have been previously published~\citep[][hereafter H21]{he21}. On this basis, we revisited the remaining candidates and confirm the genuine new OCs among them using Gaia EDR3 in this work.
 
The remainder of this this article is organized as follows. We introduce our data and methods in Section~\ref{method}, including data pre-processing, the clustering algorithm, cross-matching with other catalogs and the isochrone fitting. The results are presented in Section~\ref{results}, and a brief discussion is shown in Section~\ref{discussion}. Finally, a summary and future prospects are provided in Section~\ref{sec:summary}.

\section{Method}\label{method}

\subsection{Search and identification }
 
In our previous work (H21), we performed a blind search for OCs in Gaia DR2 which yielded 986 candidates and could confirm 74 of them.
Briefly, we adopted the DBSCAN algorithm noted earlier to discover groupings in astrometric parameter space. The principle of the algorithm is to judge whether there are at least $N$ data points within radius $\epsilon$ near each reference point. 
If the number of neighboring points is equal to or larger than $N$, the reference point is assigned as the core of a cluster, and the neighbors are deemed as cluster members, any remaining points are regarded as noise.
We selected stars with a Galactic latitude of |b| < 20$^{\circ}$, parallax $\varpi$ > 0.2~mas and G < 18~mag, so as to avoid the high density of fainter field stars, and abandon the use of stars with large uncertainties in parallax and proper motions. Then, we divided the sky into 2100 rectangles with sides of 2 $\times$ 2 degrees (|b| < 5$^{\circ}$) or 3 $\times$ 3 degrees (|b| > 5$^{\circ}$), and ran the clustering algorithm on the stars in each grid. 

It is worth mentioning that each data point contains a five-dimensional vector. Different from the input parameters $(l,b,\mu_{\alpha^*},\mu_{\delta},\varpi)$ utilized in most  studies~\citep[e.g.][]{Castro20, hunt21}, our clustering method  adopts the projected distance and linear velocity as input data for the clustering analysis as: 
\begin{equation}\label{eq1}
(d_{l^*},d_b,v_{\alpha^*},v_{\delta},\varpi)=
 (d\cdot\sin\theta_{l}\cdot\cos b,d\cdot\sin\theta_b,d\cdot\mu_{\alpha^*},d\cdot\mu_{\delta},\varpi)
\end{equation}

, where distance $d$ is taken to be the inverse of the parallax,  where $ \theta_{l} $ and $ \theta_{b} $ are the angular sizes from the star to the  center of the rectangle in $l$ and $b$, respectively.

This means, we use the stellar distance obtained by the reverse of the parallax, and take the center of each grid as the origin to calculate the projected values of the actual distance and relative velocity of each star in the grid from the origin point.
The advantage of this approach is that clustering always occurs at the same vector region in parallax, position, and velocity \textbf, and it is not sensitive to the size of the cluster radius and the dispersion of the proper motions, which change for different distances. 
Therefore, this processing can reduce the influence of the selected grid side length and avoid dense regions in the foreground and background along the line of sight. 

Additionally, we use a bimodal Gaussian function (H21) to fit k$_{th}$ nearest neighbor distance ($kNND$) for each data point $i$:
 \begin{equation}\label{eq2}
N_{kNND}=\sum_{i=1}^{2}a_i \cdot e^{\frac{-(\mathbf{d}_{k,nn}-\mu_i)^2}{2\sigma_i^2}}
\end{equation}
  \begin{equation}\label{eq3}
D_{ik}=\sqrt{(x_{i1}-x_{k1})^2+(x_{i2}-x_{k2})^2+...+(x_{i5}-x_{k5})^2}
\end{equation}
so that the $kNND$ of possible cluster members and field stars follows a Gaussian distribution with different parameters, and the intersection point is taken as the radius parameter, $\epsilon$, for DBSCAN, which is used to distinguish field stars from cluster members. 

However, it is worth mentioning that the minimum number of cluster neighbors used in the previous clustering process (H21) only takes one constant value, $N$ = 8. Therefore, the member stars sample it finds may not be complete, and the algorithm may find only parts of some clusters located at the edge of the grid described above, or just discern some stellar asterisms along the line of sight. 
Despite these caveats, in this work, we have continued to use Gaia EDR3 data to revisit the previously found OC candidates. 

The algorithm we used to estimate the membership probability of each input star is pyUPMASK \citep{pyupmask}, which is an open source software package compiled in the Python language following the development principle of Unsupervised Photometric Membership Assignment in Stellar Clusters~\citep[UPMASK,][]{upmask}, a member star determination method developed to process photometric data; however, it was later widely used in the determination of member stars based on astrometric parameters~\citep[e.g.][]{ CG19-0,cg20arm}.

The key assumptions of UPMASK include: (i) cluster members have similar properties (e.g. the clustering distribution both in proper motion and parallax) and (ii) their spatial distribution is more crowded than a random uniform distribution. The main steps of UPMASK include: (a) use the K-means clustering method to determine clumps in three-dimensional astrometric space ($\mu_l $, $\mu_b $, $\varpi$) and (b) according to the estimation results of the kernel density (KDE), determine whether these small clumps have a clustered distribution in all parameter space (compared with a random uniform distribution). If so, it is preliminarily determined that a star in a small clump is a cluster member based on its similar spatial and kinematic properties relative to other stars in the clump.

Finally, the cluster membership probability is determined as follows:
  \begin{equation}\label{eqmp}
P_{cl}=KDE_{m}/(KDE_{m}+KDE_{nm})
\end{equation}
,where $KDE_{m}$ and $KDE_{nm}$ are the KDE likelihood for the members and non-members, respectively.

Gaia EDR3 contains more five parameter astrometric data than in DR2, and the parallax and proper motion accuracy have been further improved. 
We downloaded data from Gaia EDR3 using the following rough criteria to ensure completeness and accuracy for cluster member candidates: G < 19~mag; ruwe < 1.4; $\varpi$ > 0.1~mas.
The selected samples contained 912 (986 - 74 confirmed) new OC candidates. 
For each of the candidate clusters, we selected stars within five times the standard deviation ($5 \sigma$) of their parameters ($l$, $b$, proper motion, parallax), which were also obtained in our previous work H21.
The probability of member stars is calculated with the pyUPMASK algorithm. Then, based on the membership determination results, a visual inspection of the results was performed. 
Since cluster members are considered to be gravitationally bound stars with similar motions and ages, they should exhibit clustering characteristics in position, kinematic space, and color-magnitude diagrams (CMDs).
Inspecting the characteristics of each cluster's parallax distribution (G magnitude vs parallax), proper motion VPD, CMD and spatial distribution morphology, 601 cluster candidates were identified.

\subsection{Cross matching}\label{cross}
During the course of conducting this present study, more new OCs and candidates were found and published based on Gaia DR2 data~\citep[e.g.][]{hunt21,ferreira21,casado21}. We collected these new findings and cross matched them with our candidates. At the same time, we also cross matched the new candidates with earlier catalogs, such as those of \citet[][]{CG18,CG19-0,Castro18,Castro19,Castro20,Sim19,Liu19,Ferreira19,Ferreira20,Hao20,Qin20} and H21. 
For each cross-match, an initial match was taken within the 5$\sigma$ region of the corresponding candidate's radius in Galactic longitude and latitude. Then, we compared each filtered OC with the new candidate by visual checking. Based on position, parallax, and proper motion distributions, star clusters within the same range without obvious distinction were considered to be a matched cluster, 59 candidates were removed via this process. After that, we also made a comparison with the MWSC catalog~\citep[]{Kharchenko13} and DAML02 catalog~\citep[]{Dias02} to compare the candidates’ radius, distance, and age parameters with the cataloged OCs; as a result, one more candidate was removed. Finally, 541 new candidates remained in our catalog, which were not present in the above studies. 

Recently, ~\citet[]{castro22} reported more than 600 new star cluster candidates based on Gaia EDR3 data, posted on the astro-ph archive. Since they have not given all star cluster lists at the time of we applied the final revision to this paper, we did not cross match with their samples. However, the parallaxes of the star cluster candidates obtained by \citet[]{castro22} are mostly less than 1~mas, only 4 (132 in this work) candidates within 1~kpc of the Solar System and only 79 have parallaxes lager than 0.5~mas (405 in this work).
It is our intention to cross match the OC candidates found here with the samples of \citet[]{castro22} in a future work (in prep.). 
 
\subsection{Isochrone fitting}\label{isochrone}  
Using the isochrone-fitting method with the PARSEC theoretical isochrone models \citep{Bressan12} updated by the \emph{Gaia} EDR3 passbands using the photometric calibrations from ESA/Gaia
\footnote{\url{https://www.cosmos.esa.int/web/gaia/edr3-passbands}} \footnote{\url{http://stev.oapd.inaf.it/cgi-bin/cmd_3.6}}, we derived their physical parameters. The isochrones used here contain logarithmic ages from 6.0 to 9.9 with a step of 0.1, and metal fractions from 0.002 to 0.04 with a step of 0.002. 
The isochrones were corrected for extinction and reddening, using extinction $A_G$ from 0 to 5~mag, with a step of 0.05~mag; and we adopted a coefficient value of $A_G = 1.86 E(BP-RP)$ to get the reddening values, which using ~\citet{Cardelli89} and~\citet{Donnell94} extinction curve with $R_V = 3.1$. In this process, following the method of \citet{Liu19} and H21, an automatic fitting function
\begin{equation}\label{eqage}
\bar{d^2}= \frac{\sum_{k=1}^{n}(\mathbf{x}_k-\mathbf{x}_{k, nn})^2}{n}
\end{equation}
was applied to all new cluster candidates, where $n$ is the number of selected members in a cluster candidate, and $\mathbf{x}_k$ and $\mathbf{x}_{k, nn}$ are the positions of the member stars and the points on the isochrone that are closest to the member stars, respectively. To avoid the large uncertainties of the faint stars, and reduce the influence of non-members, we restricted the isochrone fitting to stars as follows:
\begin{itemize}
\item[\textbullet] G $\leq$ 18 mag,
\item[\textbullet] $P_{cl}$ > $p$  if N ($P_{cl}$ > $p$ ) > 30 stars, for $p \in [0.8,0.9]$.
\item[\textbullet] $P_{cl}$ > $p$ if N ($P_{cl}$ > $p$ ) > 15 stars, for $p \in [0.3,0.4,0.5,0.6,0.7]$
\end{itemize}

The function outputs the parameters that are most consistent with the theoretical isochrone values. In this process, the fitting accuracy is limited by the selection of each parameter step. Using the method of~\citet{Liu19}, the uncertainties of age, extinction, distance modulus, and metallicity were estimated as half a step in the isochrone fitting, i.e., 0.05 dex for age, 0.025 for extinction and distance modulus, and 0.001 for metallicity.

\section{Results}\label{results}

 \subsection{ Classification   }

\begin{figure*}
\begin{center}
	\includegraphics[width=1.0\linewidth]{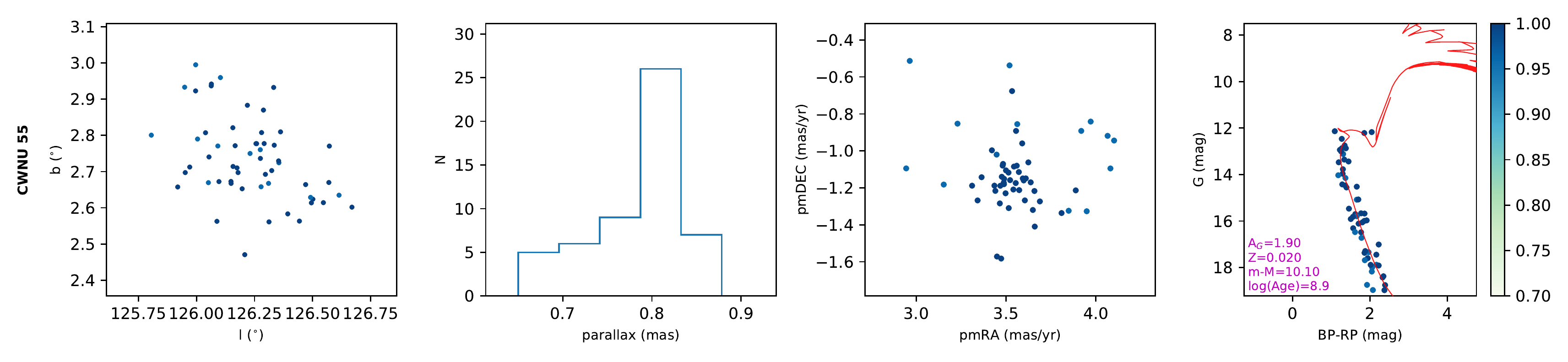}
	\includegraphics[width=1.0\linewidth]{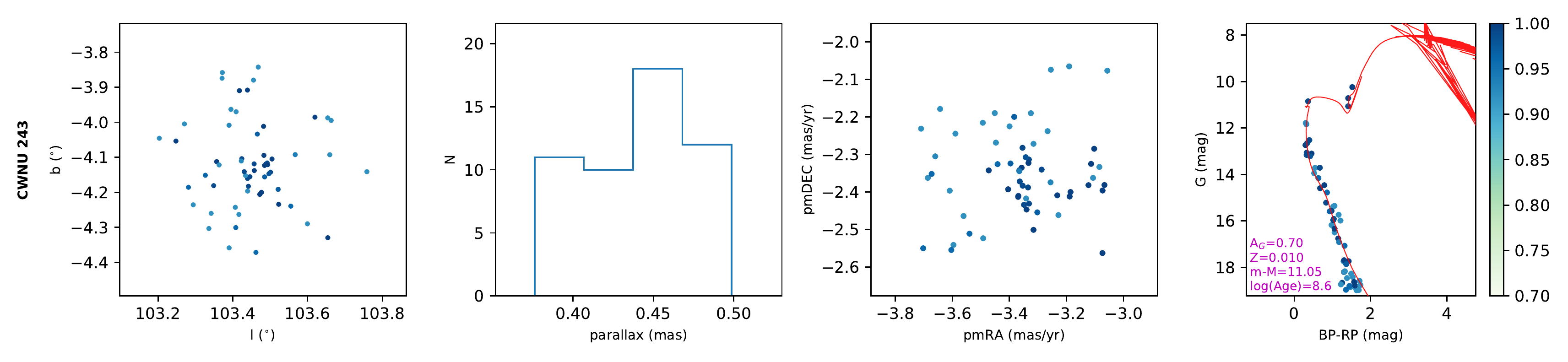}
	\includegraphics[width=1.0\linewidth]{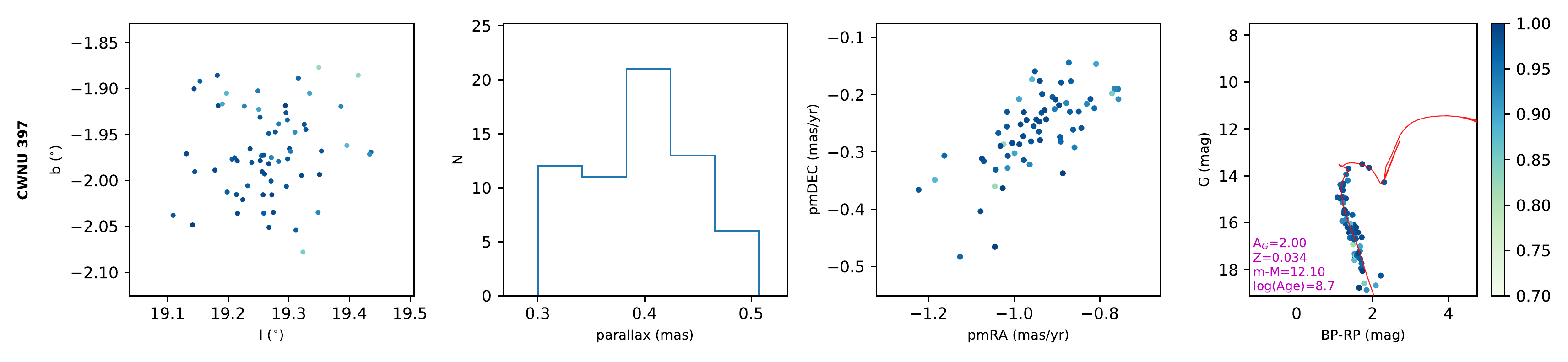}
		\includegraphics[width=1.0\linewidth]{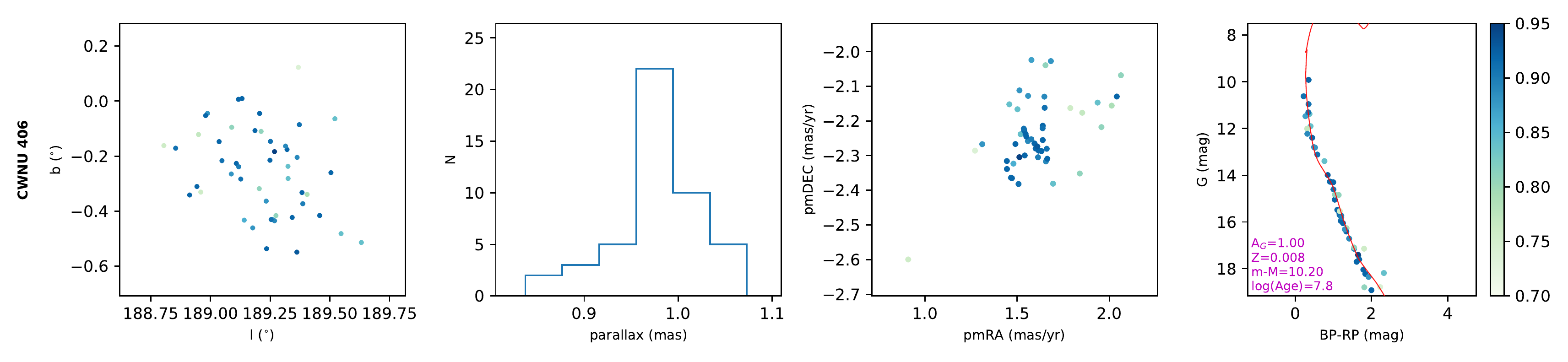}
		\includegraphics[width=1.0\linewidth]{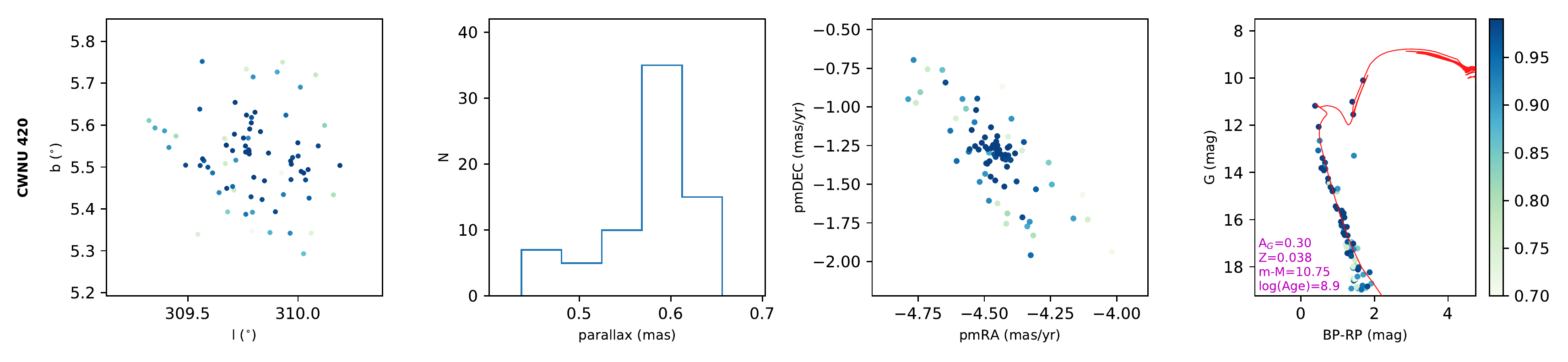}
	\caption{Astrometric and photometric examples of the new OC candidates in class 1. The color bars represent the cluster probability ($P_{cl}$) of the member stars. The complete class 1 pictures can be seen in Fig. Set~1.}
	\label{fig1}
\end{center}
\end{figure*}

\begin{figure*}
\begin{center}
	\includegraphics[width=1.0\linewidth]{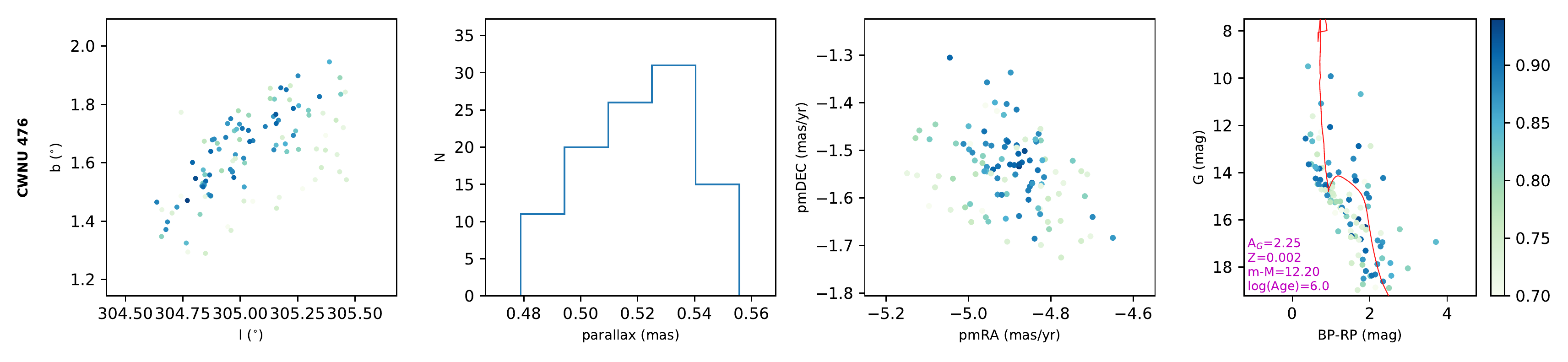}
	\caption{Same as Fig.~\ref{fig1} but for the new OC candidates in class 2. The complete class 2 pictures can be seen in Fig. Set~2.}
	\label{fig2}
\end{center}
\end{figure*}

\begin{figure*}
\begin{center}
	\includegraphics[width=1.0\linewidth]{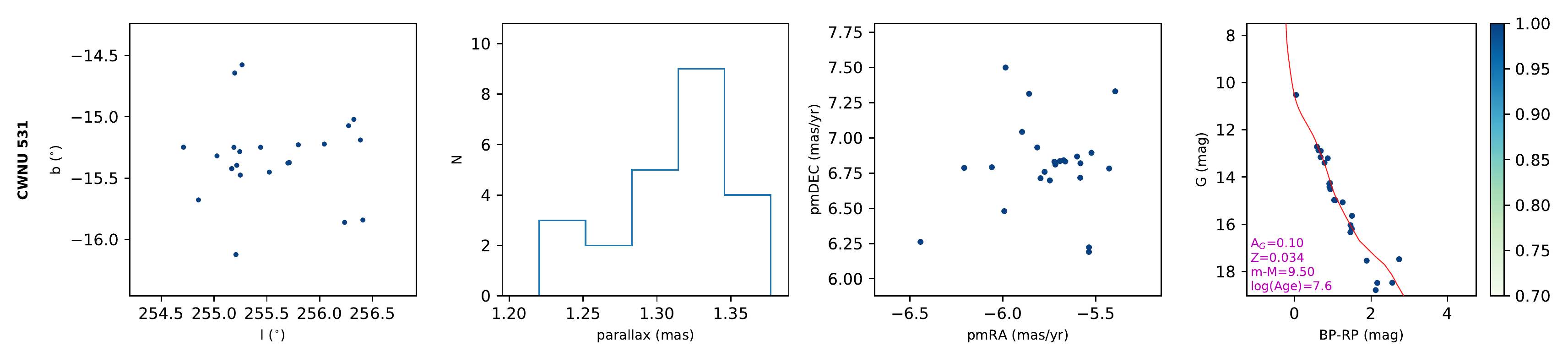}
	\caption{Same as Fig.~\ref{fig1} but for the new OC candidates in class 3. The complete class 3 pictures can be seen in Fig. Set~3.}
	\label{fig3}
\end{center}
\end{figure*}

We examined each new candidate visually, and divided them into three classes according to the CMD morphology and isochrone fitting results. The cluster candidates in class 1 have a sufficient number of member stars and clear CMDs (see examples in Fig.\ref{fig1}). Class 2 includes candidates with unclear isochrone fitting and loosely CMD distribution (see example in Fig.\ref{fig2}).
At the same time, for the two obvious failed isochrone fitting candidates, No.455 and No.493, we did the fitting manually.
 Similar to \citep{Castro20,castro22}, class 3 contains candidates with fewer stars and longer gaps in the Main Sequence (see example in Fig.\ref{fig3}). The number of cluster candidates in class 1, class 2 and class 3 is 452, 56, and 33, respectively. The complete figure sets (541 images) are available in the online journal.


The classifications are based only on CMDs and isochrone fittings, and they do not represent the distinction between real and pseudo clusters. For the cluster candidates of class 2 and 3, the CMDs are not clear enough. The possible reason is the selection of cluster member stars is still incomplete, and it may be caused by severe differential reddening and/or the passband flux contamination in dense stellar regions \citep{Evans18,Arenou18,CG18}. Besides, it should be noticed that, for many cases (especially for the new candidates with low member stars) where there are no stars at the Main Sequence turn-off, the derived ages and metallicities are less certain than for other cases.
 
\subsection{Cluster Parameters}\label{para}
We calculated the median positions, parallax, proper motions, and their corresponding dispersions~\citep[similiar to][]{CG20_0} for the new 541 candidates.  Among them there are 26 candidates with fewer than 30 member stars, and whose minimum membership probability $P_{cl}$ is 0.01. For these candidates we adopted all the member stars to do the calculation.
For the 385 candidates that possessed more than 30 member stars and with a $P_{cl}$ greater than 0.7 (N$_{70}\geqslant$ 30), we calculated statistical values for those $P_{cl} \geqslant$ 0.7 member stars. For the remaining 130 candidates, we selected different minimum membership probabilities ($min_{P_{cl}}$) to ensure that each candidate had more than 30 members. See Table 1 for examples of the median cluster parameters, as well as the number of member stars (N$_{min_{P_{cl}}}$) and the $min_{P_{cl}}$ values used in the statistical analysis.

\begin{longrotatetable}
\begin{deluxetable*}{cccccccccccccccccccccc}
	\tablecaption{Derived astrophysical parameters for the 541 cluster candidates identified in this work. The positions, parallax, and proper motions of each candidate have been calculated as the median value of selected cluster members for $P_{cl}$ > $min_{P_{cl}}$. The dispersion of each astrometric value is also presented, except in the case of $N_{RV}$ = 1, in which the RV uncertainty is given instead.}
	\label{table_all}
	\tablewidth{0pt}
    \tabletypesize{\tiny}
\tablehead{ $Cluster$ & $l$& $\sigma_{l}$ & $b$  & $\sigma_{b}$ & $Parallax$ & $\sigma_{Parallax}$ & $pmRA^{\ast}$ & $\sigma_{pmRA^{\ast}}$ & $pmDEC$ & $\sigma_{pmDEC}$  & $RV$ & $\sigma_{RV}$  & $N_{RV}$ & $m-M$ & $A_G$  & $log(age/yr)$ & $Z$ &  $N_{70}$ & $min_{P_{cl}}$ & $N_{min_{P_{cl}}}$ & $Class$\\ 
   & ($^{\circ}$)  & ($^{\circ}$) & ($^{\circ}$) & ($^{\circ}$) & (mas) & (mas) & (mas~  yr$^{-1}$) & (mas~  yr$^{-1}$) & (mas~  yr$^{-1}$) & (mas~  yr$^{-1}$)  & (\kms)  & (\kms)  &  &  & (mag)  &  &  &  &  &  & }
	\startdata
CWNU 1 & 346.147 & 0.06 & -2.132 & 0.06 & 0.60 & 0.06 & -0.03 & 0.12 & -3.58 & 0.08 & -18.1 & 0.2 & 1 & 10.60 & 2.60 & 8.3 & 0.008 & 115 & 0.70 & 115 & 1\\ 
CWNU 2 & 283.662 & 0.61 & -11.431 & 0.22 & 1.39 & 0.04 & -9.45 & 0.14 & 5.90 & 0.23 & 8.6 & 3.3 & 3 & 9.50 & 0.45 & 7.7 & 0.036 & 24 & 1.00 & 24 & 1\\ 
CWNU 3 & 131.435 & 0.20 & 3.074 & 0.19 & 1.28 & 0.03 & -2.42 & 0.09 & -1.06 & 0.09 & -7.6 & 17.8 & 2 & 9.20 & 1.45 & 7.6 & 0.006 & 48 & 0.96 & 48 & 1\\ 
CWNU 4 & 348.551 & 0.05 & -0.988 & 0.06 & 0.60 & 0.04 & 0.99 & 0.12 & -1.14 & 0.07 &   &   & 0 & 11.10 & 2.60 & 8.4 & 0.030 & 55 & 0.73 & 55 & 1\\ 
CWNU 5 & 176.325 & 0.24 & -10.221 & 0.17 & 2.27 & 0.12 & 1.27 & 0.17 & -5.95 & 0.24 & 18.9 & 8.5 & 1 & 8.75 & 1.20 & 6.8 & 0.022 & 26 & 0.40 & 30 & 1\\ 
CWNU 6 & 322.557 & 0.16 & 0.888 & 0.17 & 0.60 & 0.03 & -5.05 & 0.15 & -3.01 & 0.09 & -34.0 & 1.7 & 2 & 10.50 & 1.50 & 8.9 & 0.022 & 27 & 0.60 & 32 & 1\\ 
CWNU 7 & 325.134 & 0.17 & -4.615 & 0.05 & 0.74 & 0.06 & -2.72 & 0.18 & -1.02 & 0.17 &   &   & 0 & 10.95 & 0.65 & 8.4 & 0.040 & 47 & 0.71 & 47 & 1\\ 
CWNU 8 & 232.626 & 0.61 & -3.189 & 0.30 & 1.02 & 0.09 & -3.72 & 0.34 & 4.31 & 0.23 &   &   & 0 & 9.95 & 0.25 & 7.8 & 0.022 & 43 & 0.71 & 43 & 1\\ 
CWNU 9 & 218.107 & 0.18 & -7.662 & 0.15 & 0.99 & 0.04 & -3.38 & 0.10 & 0.68 & 0.13 &   &   & 0 & 9.80 & 1.10 & 7.0 & 0.006 & 23 & 0.50 & 41 & 1\\ 
CWNU 10 & 192.751 & 0.21 & 2.087 & 0.13 & 1.01 & 0.07 & -1.74 & 0.21 & -2.85 & 0.08 &   &   & 0 & 10.35 & 0.65 & 8.4 & 0.040 & 42 & 0.84 & 42 & 1\\ 
CWNU 11 & 191.296 & 0.16 & -1.835 & 0.15 & 1.08 & 0.06 & -0.17 & 0.11 & -1.50 & 0.09 &   &   & 0 & 9.60 & 0.95 & 7.8 & 0.004 & 32 & 0.74 & 32 & 1\\ 
CWNU 12 & 260.616 & 0.25 & -4.463 & 0.11 & 0.70 & 0.03 & -5.01 & 0.12 & 4.47 & 0.10 &   &   & 0 & 11.05 & 0.85 & 7.4 & 0.026 & 69 & 0.71 & 69 & 1\\ 
CWNU 13 & 89.188 & 0.08 & -0.303 & 0.08 & 0.49 & 0.03 & -2.53 & 0.08 & -4.01 & 0.07 & -18.4 & 0.2 & 1 & 11.45 & 1.60 & 7.6 & 0.040 & 109 & 0.70 & 109 & 1\\ 
CWNU 14 & 123.998 & 0.47 & -9.153 & 0.53 & 1.55 & 0.07 & -4.22 & 0.34 & -2.52 & 0.23 & -20.5 & 0.7 & 2 & 9.50 & 0.70 & 7.7 & 0.028 & 48 & 0.96 & 48 & 1\\ 
CWNU 15 & 182.864 & 0.27 & -0.617 & 0.17 & 0.88 & 0.05 & -0.39 & 0.22 & -2.70 & 0.25 &   &   & 0 & 9.85 & 1.20 & 7.5 & 0.002 & 29 & 0.01 & 50 & 1\\ 
CWNU 16 & 140.707 & 0.14 & -10.359 & 0.14 & 0.95 & 0.04 & 2.70 & 0.24 & -4.84 & 0.26 & -1.0 & 6.5 & 2 & 9.75 & 0.45 & 9.0 & 0.006 & 32 & 0.92 & 32 & 1\\ 
CWNU 17 & 148.794 & 0.07 & -3.200 & 0.06 & 0.33 & 0.06 & 0.80 & 0.06 & -1.25 & 0.09 &   &   & 0 & 12.40 & 2.35 & 8.9 & 0.032 & 108 & 0.70 & 108 & 1\\ 
CWNU 18 & 219.760 & 0.38 & -18.082 & 0.29 & 1.48 & 0.09 & -1.84 & 0.33 & 1.64 & 0.25 & 24.2 & 13.8 & 4 & 8.60 & 0.75 & 7.1 & 0.018 & 103 & 0.71 & 103 & 1\\ 
 ...  & ...  & ...  & ...  & ... & ...   & ...  & ...  & ...  & ...   & ...  &...  & ...  & ...  & ...  & ...  & ...  & ...  & ...  & ...  & ...  & ...  \\ 
CWNU 453 & 293.658 & 0.06 & -1.637 & 0.04 & 0.40 & 0.02 & -6.38 & 0.14 & 1.28 & 0.10 &   &   & 0 & 12.45 & 2.65 & 6.1 & 0.002 & 38 & 0.72 & 38 & 2\\ 
CWNU 454 & 122.535 & 0.03 & 1.048 & 0.04 & 0.30 & 0.02 & -2.36 & 0.06 & -0.26 & 0.12 &   &   & 0 & 13.50 & 2.55 & 6.8 & 0.038 & 30 & 0.70 & 30 & 2\\ 
CWNU 455 & 78.762 & 0.09 & 2.959 & 0.09 & 0.78 & 0.04 & -2.38 & 0.13 & -3.02 & 0.15 &   &   & 0 & 11.35 & 3.70 & 8.6 & 0.040 & 37 & 0.72 & 37 & 2\\ 
CWNU 456 & 69.761 & 0.03 & 0.808 & 0.03 & 0.46 & 0.04 & -1.06 & 0.08 & -1.98 & 0.13 &   &   & 0 & 12.85 & 3.35 & 6.8 & 0.040 & 69 & 0.73 & 69 & 2\\ 
CWNU 457 & 115.780 & 0.09 & 5.106 & 0.11 & 0.47 & 0.03 & -2.25 & 0.06 & -0.57 & 0.08 & -49.2 & 1.2 & 1 & 11.20 & 3.65 & 8.7 & 0.038 & 23 & 0.31 & 33 & 2\\ 
CWNU 458 & 321.298 & 0.08 & 1.440 & 0.05 & 0.47 & 0.03 & -1.60 & 0.08 & -2.13 & 0.10 &   &   & 0 & 12.95 & 3.70 & 6.7 & 0.030 & 24 & 0.52 & 34 & 2\\ 
CWNU 459 & 111.207 & 0.12 & 0.445 & 0.14 & 0.34 & 0.03 & -4.44 & 0.10 & -1.78 & 0.07 &   &   & 0 & 11.90 & 1.65 & 7.8 & 0.038 & 5 & 0.10 & 47 & 2\\ 
 ...  & ...  & ...  & ...  & ... & ...  & ...  & ...  & ...  & ... & ...  &...  & ...  & ...  & ...  & ...  & ...  & ...  & ...  & ...  & ...  & ...  \\ 
CWNU 509 & 250.882 & 0.35 & 12.297 & 0.47 & 1.05 & 0.03 & -6.47 & 0.15 & 2.71 & 0.07 &   &   & 0 & 10.85 & 0.20 & 6.5 & 0.002 & 31 & 0.73 & 31 & 3\\ 
CWNU 510 & 70.820 & 0.31 & 13.367 & 0.20 & 2.84 & 0.14 & 1.81 & 0.35 & -1.72 & 0.39 & -18.1 & 4.5 & 4 & 8.05 & 0.30 & 7.6 & 0.024 & 22 & 0.40 & 33 & 3\\ 
CWNU 511 & 63.314 & 0.14 & -4.323 & 0.17 & 0.68 & 0.05 & -1.60 & 0.13 & -4.74 & 0.20 &   &   & 0 & 10.95 & 2.00 & 7.6 & 0.034 & 25 & 0.61 & 34 & 3\\ 
  ...  & ...  & ...  & ...  & ... & ...  & ...  & ...  & ...  & ... & ...  &...  & ...  & ...  & ...  & ...  & ...  & ...  & ...  & ...  & ...  & ...  \\ 
CWNU 541 & 118.466 & 0.07 & 10.427 & 0.11 & 0.88 & 0.04 & -0.83 & 0.10 & -0.44 & 0.26 &   &   & 0 & 11.10 & 1.65 & 7.5 & 0.038 & 17 & 1.00 & 17 & 3\\  
	\enddata
\end{deluxetable*}
\end{longrotatetable}

\begin{longrotatetable}
\begin{deluxetable*}{ccccccccccccccc}
	\tablecaption{The astrometric and photometric parameters of the member stars from Gaia EDR3~\citep{gaia2021} and DR2~\citep[RV values,][]{Gaia18-Brown}. The P$_{cl}$ and CWNU ID are the estimated probability derived from pyUPMASK (see Secion~\ref{method}) and the corresponding cluster candidates, respectively.}
	\tablewidth{0pt}
	\label{table_mem}
    \tabletypesize{\tiny}
\tablehead{ source$\_$id&	l	&b	&parallax&	parallax$\_$error	&pmra	&pmra$\_$error&	pmdec&	pmdec$\_$error	&phot$\_$g$\_$mean$\_$mag	&bp$\_$rp&	dr2$\_$radial$\_$velocity	&dr2$\_$radial$\_$velocity$\_$error	&P$_{cl}$&CWNU ID\\ 
   & ($^{\circ}$)  & ($^{\circ}$)  & (mas) & (mas) & (mas~yr$^{-1}$) & (mas~yr$^{-1}$) & (mas~yr$^{-1}$) & (mas~yr$^{-1}$)  &(mag)&(mag)& (\kms)  & (\kms)  &  & }
	\startdata
5971809664655388800 & 346.359048 & -1.780253 & 0.562797 & 0.093658 & -0.250053 & 0.115445 & -3.381059 & 0.092754 & 17.387 & 2.202 &      &      & 0.01 & 1\\ 
5971775442361449984 & 345.995177 & -1.874682 & 0.642936 & 0.058110 & -0.112466 & 0.077229 & -3.605401 & 0.053768 & 16.632 & 1.926 &      &      & 0.16 & 1\\ 
5971773793094024064 & 345.933496 & -1.878416 & 0.474061 & 0.186183 & -0.037324 & 0.229708 & -3.354179 & 0.169602 & 18.564 & 2.247 &      &      & 0.01 & 1\\  
... & ... & ... & ... & ... & ... & ... & ... & ... & ... & ... & ...     & ...     & ... & ...\\ 
5959775582937720448 & 346.159082 & -2.271118 & 0.657973 & 0.020765 & -0.169403 & 0.024773 & -3.560266 & 0.019552 & 10.811 & 2.488 & -18.100 & 0.236 & 0.99 & 1\\ 
5959775548568120960 & 346.172181 & -2.288183 & 0.538414 & 0.167751 & -0.367949 & 0.209876 & -3.860625 & 0.153564 & 18.399 & 2.349 &      &      & 0.19 & 1\\ 
5959776368884700288 & 346.182605 & -2.255861 & 0.478058 & 0.090079 & -0.158356 & 0.116090 & -3.035572 & 0.094442 & 17.393 & 2.056 &      &      & 0.04 & 1\\ 
... & ... & ... & ... & ... & ... & ... & ... & ... & ... & ... & ...     & ...     & ... & ...\\ 
5248438868797049728 & 283.632602 & -10.960276 & 1.454196 & 0.009982 & -8.598344 & 0.011141 & 6.112619 & 0.012072 & 12.133 & 0.744 & 10.822 & 1.983 & 1.00 & 2\\ 
5247228173357046272 & 284.529914 & -11.841456 & 1.389711 & 0.029918 & -9.511065 & 0.038090 & 5.742784 & 0.041785 & 16.010 & 1.394 &      &      & 1.00 & 2\\ 
5248457289918918144 & 283.475789 & -11.486291 & 1.437461 & 0.030349 & -9.434943 & 0.038414 & 5.895811 & 0.039077 & 15.818 & 1.335 &      &      & 1.00 & 2\\ 
5248455876867549056 & 283.533803 & -11.345112 & 1.411003 & 0.010096 & -9.335052 & 0.014530 & 6.028354 & 0.012739 & 12.193 & 0.559 & 8.616 & 7.836 & 1.00 & 2\\ 
... & ... & ... & ... & ... & ... & ... & ... & ... & ... & ... & ...     & ...     & ... & ...

	\enddata
\end{deluxetable*}
\end{longrotatetable}
The labels of the new star cluster candidates are CWNU 1 to CWNU 541, where the abbreviation CWNU stands for China West Normal University. A full table of the parameters and the member stars along with probabilities for each new candidate (see examples in Table~2) are shown in electronic files \footnote{\url{https://cdsarc.u-strasbg.fr/ftp/vizier.submit//He22OCC/}}. 
 
Among the member stars presented above, a total of 226 candidates have Gaia DR2 radial velocities (RVs), although we note that a single star may impart a certain degree of bias to any RV evaluation. In those candidates, 111 of them have two or more member stars with RV values, and 67 ones have dispersions less than 5 \kms (see Table~\ref{tabrv}). We have also presented the median and dispersion RV values in Table 1, for candidates with $N_{RV}$ = 1, the table also shows the RV and its uncertainty for each star cataloged in Gaia DR2.
The upcominng Gaia DR3 will brings more than four times the RV numbers in Gaia DR2, the kinematic samples of star clusters will be greatly improved.

\begin{deluxetable}{ccccc}
\caption{Radial velocities.}
\label{tabrv}
\tablewidth{1pt}
\tablehead{\colhead{ } &  \colhead{$N_{RV}$ = 1}&  \colhead{$N_{RV}$ = 2}&  \colhead{$N_{RV}$ = 3}&\colhead{$N_{RV} \geqslant$ 4}}
\startdata
$\sigma_{RV}$ < 5 \kms	   & &		26 	 &	19&	22	\\
$\sigma_{RV}$ > 5 \kms    &     &      29 &	2&	13  \\
Total & 	115 &  55 & 21&	35 	\\
\enddata
\end{deluxetable}

\section{Discussion}\label{discussion}
\subsection{Comparison with other works} \label{comparison}

\begin{figure*}
\begin{center}
	\includegraphics[width=1.\linewidth]{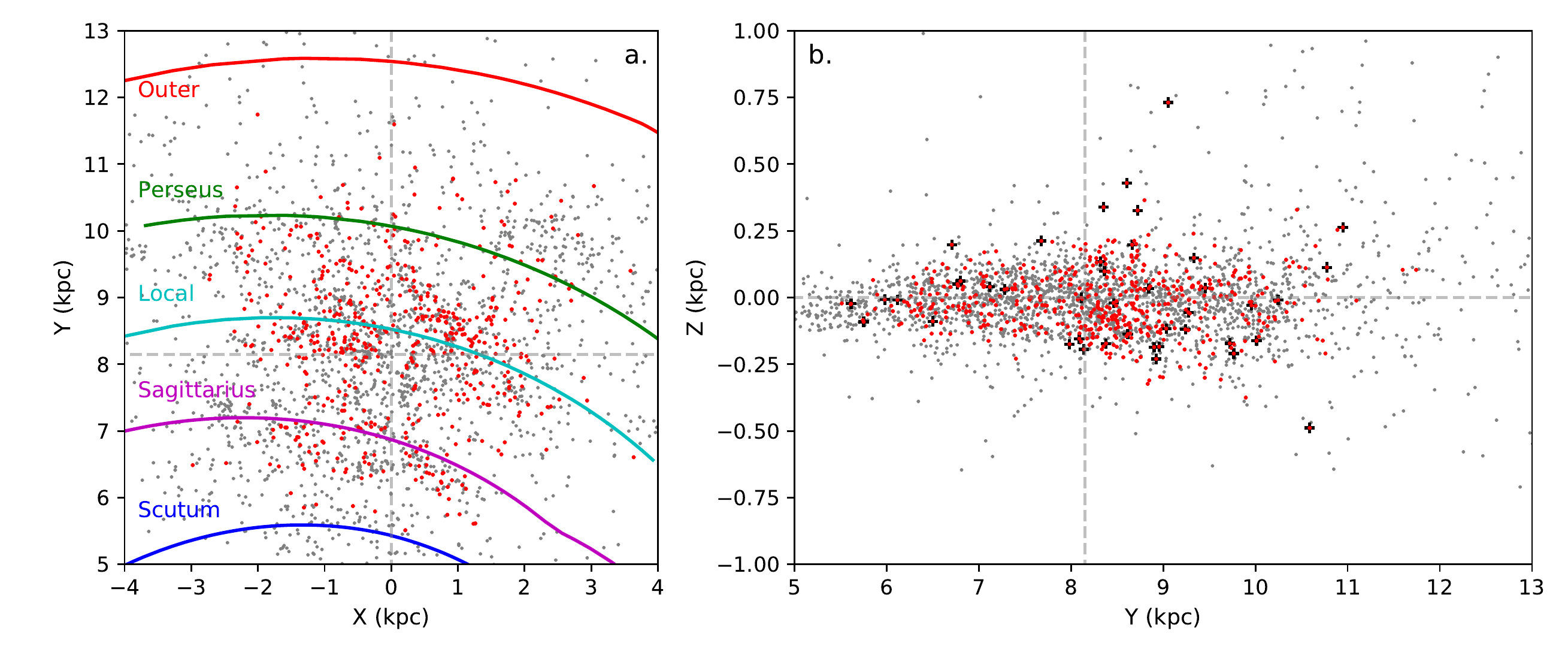}
	\caption{Locations of the new OC candidates (red dots) and known OCs (grey dots) determined by the median parallax of member stars. (a) Image viewed from the north Galactic pole from which the Galaxy rotates clockwise. The colored lines show the spiral arms determined by ~\citet{Reid19}. (b) Same as (a) but with an edge-on view. The black plus signs represent new candidates older than 10$^9$ yr. Here, we have assumed the ~\citet{Reid19} recommended values of $R_0$ = 8.15~kpc and $Z_0$ = 5.5~pc.}
	\label{fig4}
\end{center}
\end{figure*}

Most of the new cluster candidates found in this work are within 3 kpc of the Solar System. As can be seen from the Fig.~\ref{fig4}(a), the new candidates are located in the Local, Sagittarius, and Perseus arms and the inter-arm regions. The majority of them are within about 250 pc of the Galactic disk, and objects farther than 250 pc are mostly old clusters (Fig.~\ref{fig4}(b)).
 
\begin{figure*}
\begin{center}
	\includegraphics[width=0.9\linewidth]{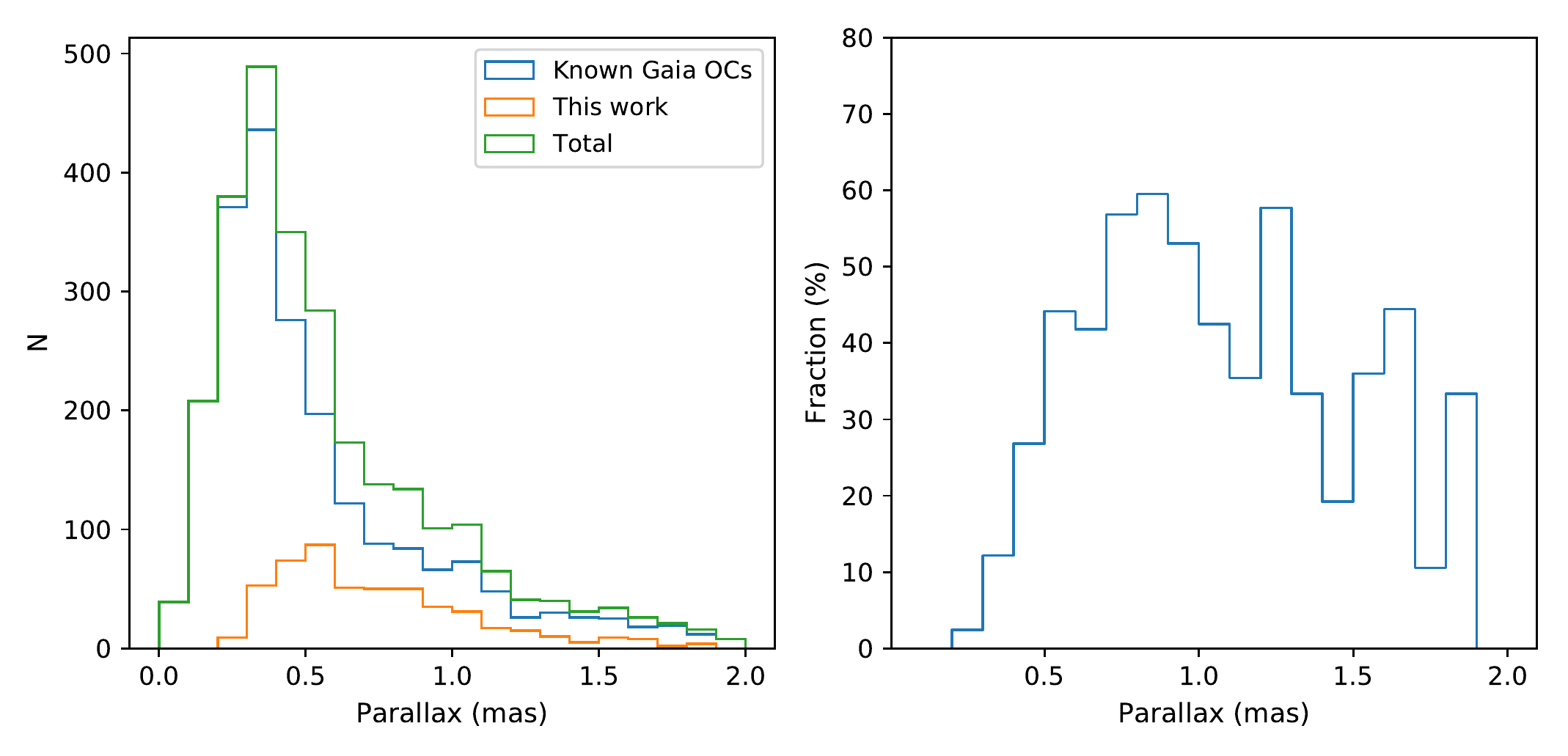}
	\caption{ Left: histograms of the parallaxes of cataloged Gaia DR2 OCs \citep[including those of][]{CG18,CG19-0,Castro18,Castro19,Castro20,Sim19,Liu19,Ferreira19,Ferreira20,ferreira21,Qin20,he21,hunt21,casado21} clusters and new OC candidates, as well as the result of adding them together. Right: proportion of increase in the number of Gaia DR2 OCs at different parallaxes.}
	\label{fig6}
\end{center}
\end{figure*}

\begin{figure*}
\begin{center}
	\includegraphics[width=0.45\linewidth]{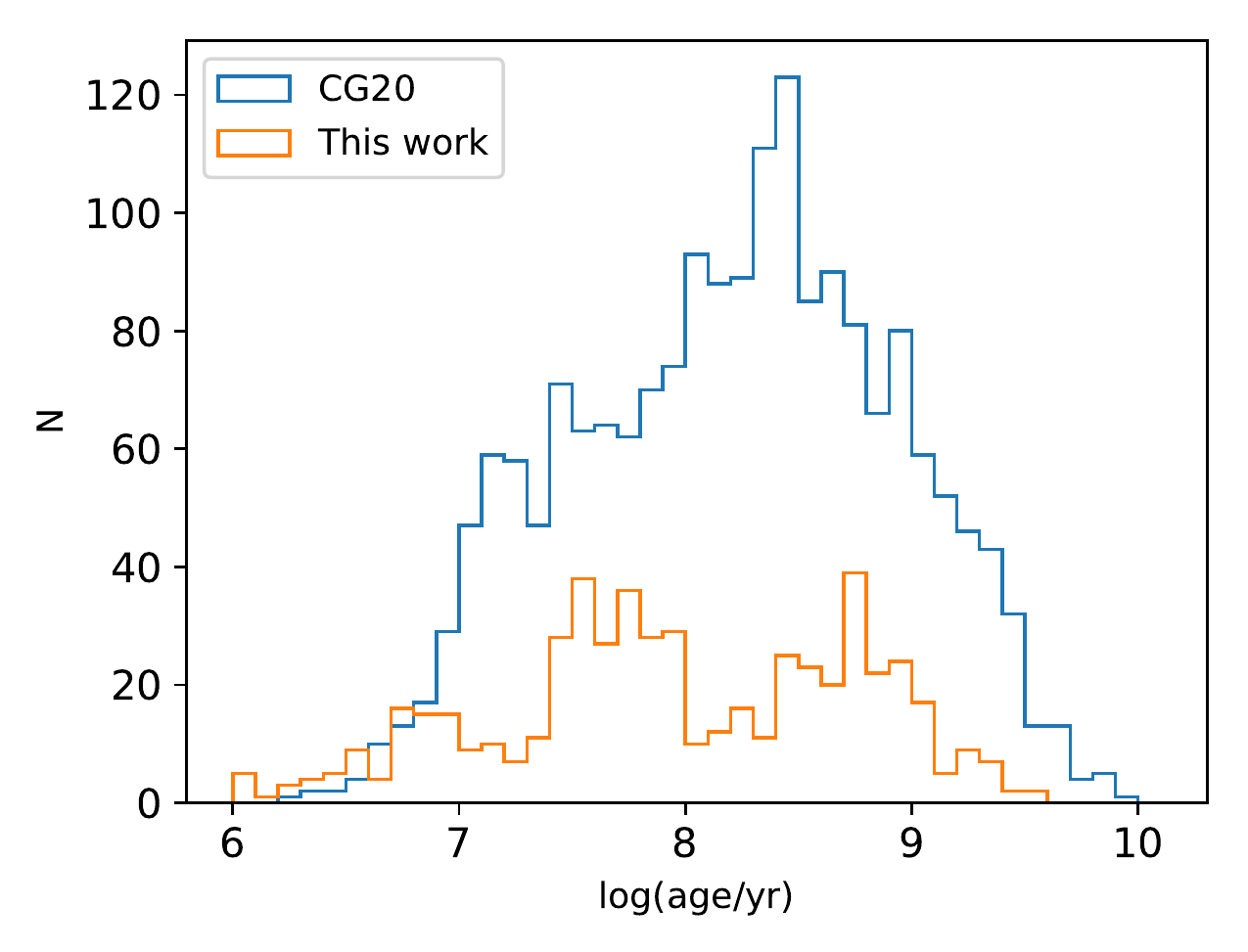}
	\includegraphics[width=0.45\linewidth]{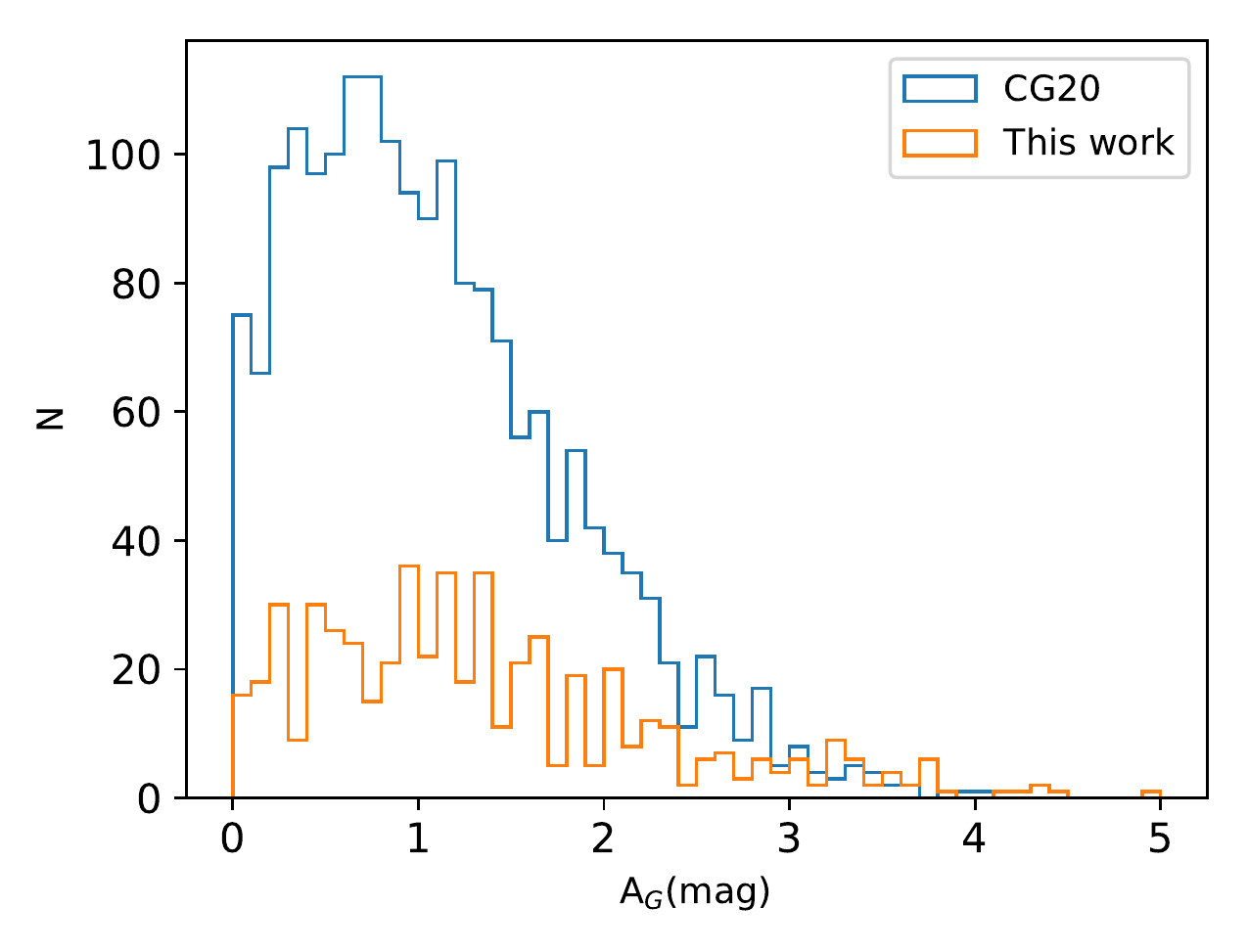}
	\caption{Histograms of the cluster ages (left), G-band extinctions (right) in CG20, and the values derived from isochrone fittings in this work.}
	\label{fig7}
\end{center}
\end{figure*}

We compared the newly discovered cluster candidates with known clusters based on Gaia DR2. As shown in Fig.~\ref{fig6}, the parallaxes of the new candidates are mostly between 0.3 and 1.5~mas.The number of the candidates found in the 0.7-0.9 mas range has increased by over $50\%$.
At the same time, the ages of the new candidates have two peaks, at 7.4-8.0 and 8.3-8.9 in log (age/yr), respectively (Fig.~\ref{fig7}, left). The G-band extinctions in the regions of the new candidates are mostly lower than 2~mag (Fig.~\ref{fig7}, right), ~which is similar to the literature~\citep[][hereafter CG20]{cg20arm}.


\subsection{ Proper motion dispersion}
As mentioned by \citet{CG20_0} and \citet{dias21}, cluster members present smaller velocity dispersions than field stars, a consequence of their gravitational bounding state. Therefore, measurements of the dispersions of the member star proper motions is a necessary (but not sufficient) condition to measure whether a stellar aggregate has the characteristics of a real cluster.
 
\begin{figure*}
\begin{center}
	\includegraphics[width=0.9\linewidth]{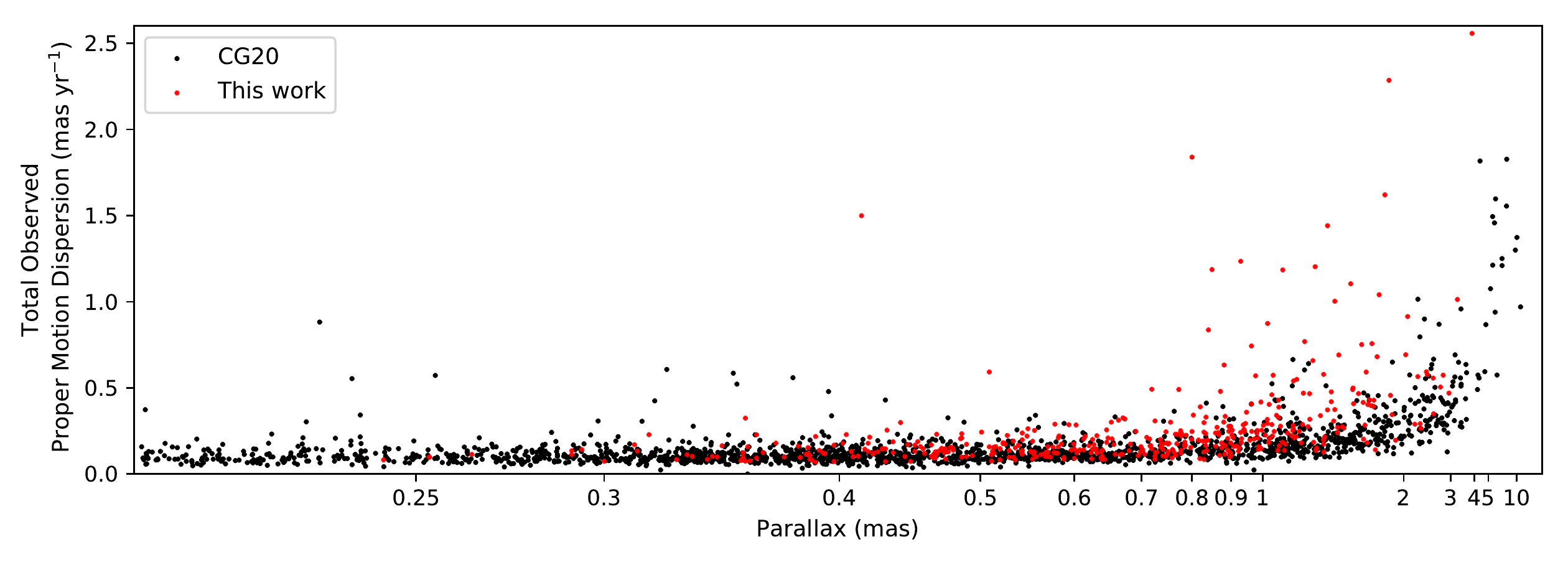}
	\caption{Total observed proper motion dispersion using Gaia EDR3 vs. parallax. The $x$-axis is a log scale for known OCs from the homogeneous OC catalog of CG20 (black dots) compared with the new candidates in this work (red dots). }
	\label{fig5}
\end{center}
\end{figure*}

We also compared the proper motion dispersions of the new OC candidates with those of CG20. Here, we used the Gaia EDR3 astrometric parameters of the OC member stars found by CG20. 
As shown in Fig.~\ref{fig5}, most of the new OC candidates have similar dispersions to previous known clusters, among which about four percent (20-25) of new OC candidates' dispersion are far away from the known OCs. From this result, we can see that the cluster candidates we found may have similar dynamics to those of known OCs, and they tend to be real physical clusters rather than asterisms.

\subsection{Metallicity gradient}

\begin{figure*}
\begin{center}
	\includegraphics[width=0.8\linewidth]{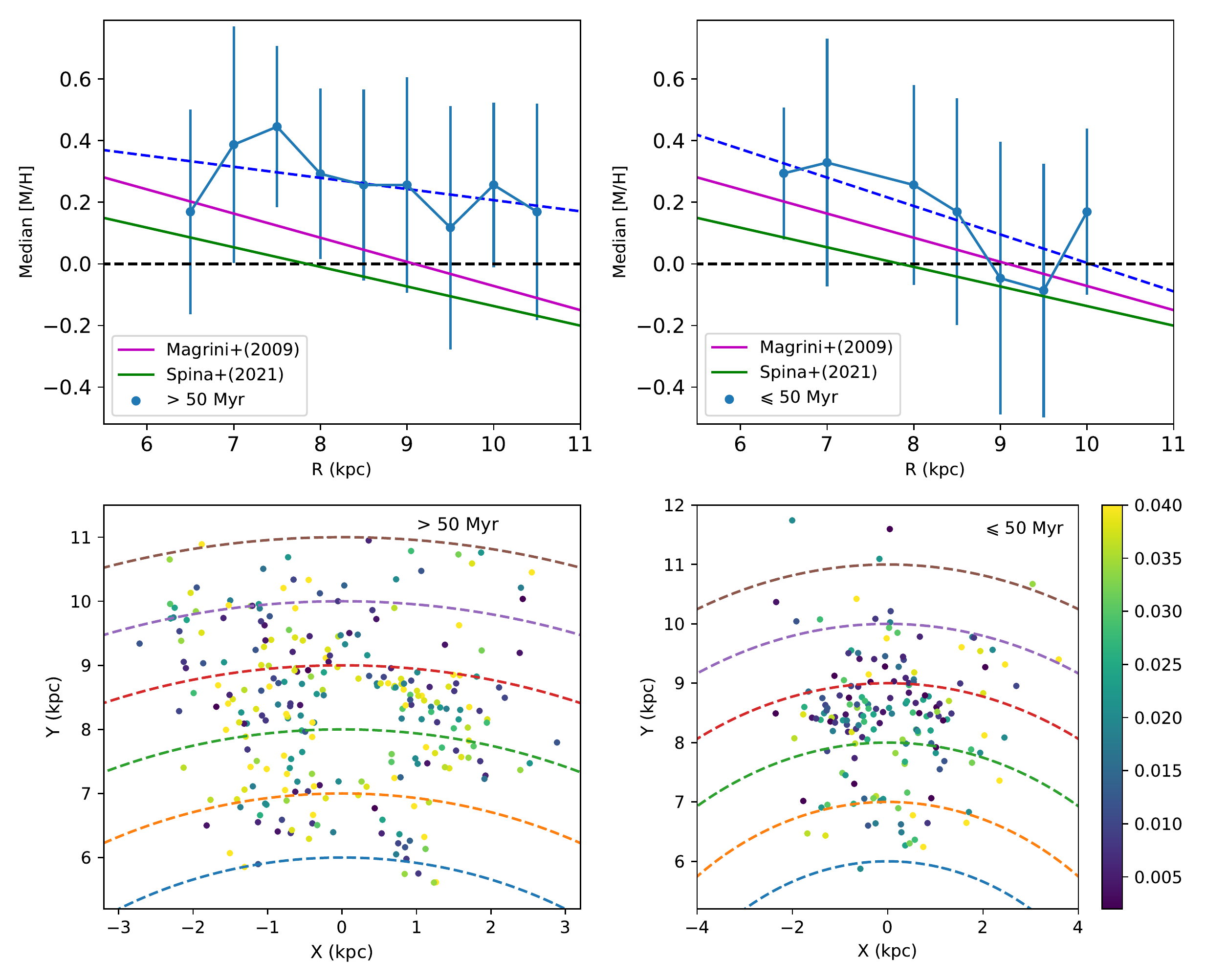}
	\caption{Metallicity  gradient~\citep[upper panels, taking solar metallicity from][]{Asplund05} for the new OC candidates and their distribution within the Galactic plane. In the lower panels, the colored dots denote their initial metal fraction. The panels show metallicity for different age ranges, left: > 50~Myr, right: $\leqslant$ 50~Myr. The blue dashed lines in upper panels show the liner fittings for median [M/H] values, the error bars show the standard deviation for [M/H] values in the corresponding radius bin ($\geqslant$ 10 candidates in each bin), and colored lines demonstrate the metallicity gradients from the literature~\citep[][]{Magrini09,Spina21}. The dashed lines in the lower panels portray the Galactocentric radius, from 6 to 11~kpc, with an interval of 1~kpc.}
	\label{fig_zini}
\end{center}
\end{figure*}

In Fig.~\ref{fig_zini}, we show the metallicity distribution of the new class 1 candidates. We have distinguished the new candidates according to two age ranges, and given the statistical results of the median metallicity values in $\Delta_{R}$ = 0.5~kpc steps along the Galactocentric radius; as before, their distribution in the Galactic plane is also shown. For older cluster candidates (> 50~Myr), the metallicity gradient is relatively moderate with radius, but a downward trend can still be seen, the metallicity gradient is about -0.036~dex kpc$^{-1}$, which is smaller than that found in the literature ~\citep[e.g. -0.076~dex kpc$^{-1}$,][]{Spina21}. 
However, the samples used here are not complete, and the metallicity dispersion values are very large, so more research is needed for further confirmations.

For the cluster candidates younger than 50~Myr, the metallicity gradient is about -0.092~dex kpc$^{-1}$, which is comparable with the predictions of previous studies~\citep[][]{Magrini09,Spina21}. More interesting, there is a suspicious steeper decrease near R $\sim$ 9~kpc, and here the younger candidates exhibit a lower metallicity distribution than older ones; this is also consistent with recent spectroscopic observations of OCs~\citep{Spina17,Spina21,Spina22}.
However, it should be noticed that these new samples are mainly located on the Local, Sagittarius, and Perseus arms near the Solar System, and whether the gradient has a wider applicability needs more relevant research using larger OC samples.

\section{Summary}\label{sec:summary}
In this paper we reported the finding of 541 new OC candidates based on our previous search in Gaia DR2 and made the new confirmations using Gaia EDR3. We applied an open source procedure pyUPMASK, to determine the probabilities of the cluster members. We provided the membership and median parameters for the new candidates, and also made classifications based on the CMDs and isochrone fittings. Our class 1 candidates presented the clear features in the CMDs, while the class 2 and class 3 candidates need more investigations regarding their member stars. 

Most of the new candidates show small total proper motion dispersions, which suggest that they may be genuine star clusters. 
 Meanwhile, the metallicity gradient of the new candidates is similar to that found in the literature; i.e., candidates in the inner disk present higher metallicity than those in the outskirts. Besides, for the youngest cluster candidates (< 50~Myr), the metallicity gradient displays a larger decrease in the vicinity of the Solar System, which is consistent with the results of earlier spectroscopic studies.

The results show that our work has significantly increased the OC sample size within 2~kpc from the Sun, Moreover, it can be seen that the cluster samples in the Gaia data are still incomplete. As a next step, we plan to explore new star clusters in the Gaia DR3 database, which will greatly increase the number of RV values for bright stars, and may assist the new generation of stellar cluster kinematics.

\section{Acknowledgements}
This work is supported by Fundamental Research Funds of China West Normal University (CWNU, No.21E030), the Innovation Team Funds of CWNU, and the Sichuan Youth Science and Technology Innovation Research Team ( 21CXTD0038). We acknowledge the science research grants from the China Manned Space Project with NO. CMS-CSST-2021-B03. Li Chen \& Jing Zhong acknowledge the support from the National Natural Science Foundation of China (NSFC) through the grants  12090040 and 12090042. Jing Zhong would like to acknowledge the NSFC under grants 12073060, and the Youth Innovation Promotion Association CAS.
This work has made use of data from the European Space Agency (ESA) mission GAIA (\url{https://www.cosmos.esa.int/gaia}), processed by the GAIA Data Processing and Analysis Consortium (DPAC,\url{https://www.cosmos.esa.int/web/gaia/dpac/consortium}). Funding for the DPAC has been provided by national institutions, in particular the institutions participating in the GAIA Multilateral Agreement.

\bibliographystyle{aasjournal} 
\bibliography{oc_v2} 


\end{CJK*}

\end{document}